\documentclass[floatfix,10pt,twocolumn,showpacs,preprintnumbers,amsmath,amssymb]{revtex4}

\usepackage{epsf}
\usepackage{graphicx}  % Include figure files
\usepackage{dcolumn}   % Align table columns on decimal point
\usepackage{bm}        % bold math

\newcommand{\be}{\begin{equation}}
\newcommand{\en}{\end{equation}}
 \newcommand{\bea}{\begin{eqnarray}}
 \newcommand{\ena}{\end{eqnarray}}

\begin{document}

\title{MOND cosmology from entropic force}
\author{Hongsheng Zhang\footnote{Electronic address: hongsheng@shnu.edu.cn} }
\author{Xin-Zhou Li \footnote{Electronic address: kychz@shnu.edu.cn} }
 \affiliation{ Center for
Astrophysics , Shanghai Normal University, 100 Guilin Road,
Shanghai 200234,China}

\date{ \today}

\begin{abstract}
 We derive the MOND cosmology which is uniquely corresponding to the original MOND at galaxy scales via entropic gravity method.
 It inherits the key merit of MOND, that is, it reduces the baryonic matter and non-baryonic dark matter into baryonic matter only.
  For the first time we obtain the critical parameter in MOND, i.e., the transition acceleration $a_c$ at cosmological scale.
   We thus solve the long-standing coincidence problem $a_c\sim cH_{0}$.   More interestingly, a term like age-graphic dark energy emerges naturally.
 In the frame of this MOND cosmology, we only need baryonic matter to describe both dark matter and dark energy in standard cosmology.

\end{abstract}


\preprint{arXiv: }
 \maketitle

\section{Introduction}

  Dark matter has been believed to be responsible for the flat rotation curves of galaxies for many years \cite{darkmatter}. Further study shows
  that dark matters share almost the same distribution in all the observed galaxies, which is independent on the shape and compositions of the galaxies. A nice analogy of this situation
  is the equivalence principle, which says that all the objects share the same acceleration, which is independent on the stuff of the object. More importantly, we never directly observe the dark matter particles. We only have some indirect evidences from their gravitational effects. Thus, a natural idea is that the surplus gravitational effect is the property of the gravity itself rather than some mysterious invisible matter. In fact, it was suggested that Newtonian dynamics is modified through replacing kinematic $a$ by an effective acceleration $a_N=a\mu(a/a_c)$, where $\mu=1$ for the usual-valued accelerations but $\mu=a/a_c$ when acceleration is much smaller than a critical acceleration $a_c$ \cite{mil1}. $a_c$ is a new scale with the same dimension of $a$. For more references about this modified Newtonian dynamics (MOND), see recent reviews\cite{revmond}.

  MOND presents excellent explanations of the rotation curves of very different types of galaxies with the same acceleration transition parameter $a_c$. This success is not surprising since MOND is designed to do so. Moreover, MOND can explain several other independent observations. For instance, Tully-Fisher relation \cite{tully} is almost axiomatic in MOND paradigm, but needs schemed distribution in dark matter scenario. We know MOND must be improved though it has splendid achievement at the galaxy scale, since it could not describe the dynamics of the universe. A systematic study of spiral galaxies pins $a_c$ to be: $a_c=1.2\times 10^{-8}{\rm cm ~sec^{-2}}=0.176cH_0$, where  $c$ is the luminous velocity, and $H_0$ denotes the present Hubble parameter.  This is a remarkable indication that MOND should be extended to cosmic scale: $a_c$ is at the scale of the present Hubble parameter $a_c\sim cH_0$, which was pointed out several years before \cite{mc}.

  A number of relativistic MOND theories appear recently. A leading relativistic MOND is tensor-vector-scalar (TeVeS) formulation \cite{beken1}.  Generally there are two scalars, one vector, and one tensor together with one arbitrary function in TeVeS, which make the theory become very complicated. The TeVeS frame involves an arbitrary
  function in addition to many parameters as well as auxiliary fields, thus it is difficult to make a testable prediction in cosmology, though we can get $\Lambda$CDM-like cosmology by fixing  some special parameter \cite{hao1}. Surely, we should reduce freedoms in this model. The best case is a theory which is uniquely corresponding to the original MOND without additional fields and arbitrary function. We shall try to find such a theory by holographic approach of gravity (entropic force, EF). Further, we shall show that such an extrapolation of MOND to the universe is well consistent with observations.

   An interesting derivation of Newtonian law and Einstein field equation is considered in \cite{ver}, in which space is emergent in holographic approach
  , and gravity  emerges when the entropy associated with the position of an object changes. In cosmology, the Friedmann equation also can be derived by EF method \cite{shu,cai1}. EF is a general paradigm which has been used to modified gravity in several aspects. For example, we can change the equipartition law with Debye distribution to get a cosmological model in which the universe accelerate without invoking dark energy \cite{gao}. FRW cosmology in Horava-Lifshitz gravity from entropic force has been introduced in \cite{lyx}. The entropic gravity in brane world scenario is explored in \cite{lingyi}. The entanglement entropic force is suggested and applied in entanglement entropy model for two black holes in \cite{chiba}.    In this letter, we prove that there exists a unique cosmology as extension of the original MOND at galaxy scale. We shall show that it inherits the critical property of MOND (replacing dark matter), and bears with an innate term to accelerate the late universe, which has never been mined in MOND before.

  This paper is organized as follows: In the next section, we derive the Friedmann equation of MOND cosmology by applying EF method to an FRW universe, based on which we explore the primary properties of MOND cosmology. In section III, we conclude this letter.

 \section{MOND cosmology}

 It is not a completely new idea that gravity acts as a mean field something like water in hydrodynamics rather than a fundamental interaction in nature \cite{saha}. It is shown that Einstein field equation can be regarded as state equation in view of thermodynamics \cite{jaco}. Recently, it is found that Einstein field equation is equal to first law of thermodynamics in spherically static symmetric spacetime \cite{pad}. It is especially important that one studies the relation between thermodynamics and gravity can not be obtained from the first principle \cite{feng1}. Now we develop the method in \cite{cai1} to deduce the Friedmann equation in MOND cosmology.

 We start from an homogeneous and isotropic 4-dimensional universe, which is described by an FRW metric,

 \be
 ds^2=-dt^2+R^2d\Omega_3^2,
 \en

 where $R$ denotes the scale factor, and $\Omega_3$ represents a 3-Euclidean space, 3-sphere, or 3-pseudosphere. Following \cite{ver,cai1}, we consider a comoving spatial spherical 3-volume $V$ with a 2-boundary $S$ as the holographic screen of $V$, which loads complete information in $V$. We introduce the prime assumption in EF approach, which says the number of bits associated to the screen $S$ is,
 \be
 N=\frac{A}{G},
 \label{num}
 \en
 where $A$ denotes the area of $S$, and as usual $G$ is the Newton gravitational constant. Note that the above equation is different from the Bekenstein-Hawking entropy in usual black hole thermodynamics due to a factor 1/4.  In the previous thermodynamical deduction of Einstein equation \cite{jaco}, the factor 1/4 is also necessary. However here, in EF approach, we must assume $N$ is directly equal to the area to keep all the other equations (3), (4), (5), (11) in their usual form and with the standard constants. This point deserves to get more attentions. Warranted by the equipartition law of energy the total energy of the screen $E$ reads,
 \be
 E=\frac{1}{2}Nk_BT,
 \label{equpar}
 \en
 where $T$ is the temperature of the comoving observer sensed by static observer (whose physical coordinates are constants ) in this dynamical FRW universe, $k_B$ is the Boltzman's constant. Similar to the arguments in \cite{ver}, the mass $M$ enclosed in $V$ equals to $E$,
 \be
 E=M.
 \label{mass}
 \en
 The temperature of the comoving observer with acceleration $a$ relative to observers with constant physical coordinates can be obtained from Unruh's reasoning \cite{unru},
 \be
 T=\frac{a}{2\pi k_B}.
  \label{tem}
 \en
 The temperature around a black hole in entropic force approach is studied in \cite{lyx2}.
 The value of $a$ can be derived directly \cite{cai1},
 \be
 a=-\ddot{R}r,
 \label{physicR}
 \en
 where $r$ is the radial comoving coordinate of the comoving observer, hence its physical coordinate $r_{ph}$ is $r_{ph}=Rr$. Here $r\in(0,1)$ from the Local Group to the boundary of the observed universe.

 Up to now, everything is standard. Then we introduce the essence of MOND: the modified acceleration. An object with acceleration $a$ is treated as an object with an effective acceleration $a_N$
 \be
 a_N=a\mu(a/a_c),
 \label{an}
 \en
  in MOND paradigm,
 where $\mu$ is a smooth function with properties that $\mu=1$ when $a/a_c>>1$ and $\mu=a/a_c$ when $a/a_c<<1$.  Then correspondingly, we use $a_N$ to replace $a$ in the Unruh formula (\ref{tem}).  Under this situation the Unruh temperature becomes,
 \be
 T=\frac{a\mu}{2\pi k_B}.
 \label{tem1}
 \en

 Next we consider a matter source in perfect fluid form,
 \be
 T_{\mu\nu}=(\rho +p) u_{\mu} u_{\nu} +pg_{\mu\nu},
 \label{em}
 \en
 in which $\rho$ and $p$ denote the density and pressure measured by comoving observers.
 The Bianchi identity requires,
 \be
 \dot{\rho}+3H(\rho+p)=0,
 \label{conti}
 \en
 where $H$ stands for Hubble parameter. We calculate $M$ enclosed in the boundary $S$ for the energy-momentum (\ref{em}). The mass which yields acceleration of the comoving observer is the active (Tolman-Komar) mass,
 \be
 M= 2\int_{ V}dV \left[ T_{\mu\nu}-\frac{1}{2}{\rm Tr}(T_{\mu\nu})
g_{\mu\nu}\right] u^{\mu}u^{\nu},
 \label{tolkom}
 \en
 where $V$ is the volume enclosed by $S$, and ${\rm Tr}(T_{\mu\nu})$ denotes the trace of the energy-momentum $T_{\mu\nu}$.
 It is straitforward to obtain the acceleration by associating the above equations (\ref{num})-(\ref{tem}), and (\ref{tolkom}),
 \be
 \frac{\ddot{R}}{R}\mu=-\frac{4\pi G}{3}(\rho+3p).
 \label{acce}
 \en
 It is clear that it just degenerates to the familiar form when $\mu=1$ ($a/a_c>>1$).

 From (\ref{acce}) and (\ref{conti}), we derive the Friedmann equation in MOND,

 \be
 H^2+\frac{k}{R^2}=\frac{8\pi G}{3\mu}\rho+\frac{1}{R^2\mu}\int dt\dot{R}^2\dot{\mu},
 \label{fried}
 \en
 where $k=1,0,-1$, depending on the spatial curvature. The first term of the RHS of (\ref{fried}) denotes the effects of baryonic matter (the total dust matter in standard model), and the second term indicates the cosmic acceleration (can happen in the early universe or late universe).

 In the present letter, we concentrate on the late time universe in MOND cosmology. To explain the observed cosmic evolution,  we introduce the concept ``virtual dark matter" and ``virtual dark energy" in MOND cosmology \cite{reviewcross}, since almost all observations are fitted in $\Lambda$CDM, or its slight variation Scalar-CDM frame.
   The generic Friedmann equation in the
 4-dimensional general relativity can be written as
 \be
 H^2+\frac{k}{R^2}=\frac{8\pi G}{3} (\rho_{by}+\rho_{dm}+\rho_{de}),
 \label{genericF}
 \en
 where the first, second, and third terms of RHS in the above equation represent the baryonic matter, dark matter, and dark energy respectively.
 Comparing (\ref{genericF}) with (\ref{fried}), we derive the virtual dark matter and dark energy in MOND cosmology,
 \be
 \rho_{dm}=\left(\frac{1}{\mu}-1\right)\rho,
 \label{rhodm}
 \en
 and
 \be
 \rho_{de}=\frac{3}{8\pi G}\left(\frac{1}{R^2\mu}\int dt\dot{R}^2\dot{\mu}\right).
 \label{vde}
 \en
 Now it is easy to derive $\mu_0$ by cosmological observations. The subscript 0 labels the present value of a quantity.  Introducing a dimensionless parameter arising naturally in  $\Lambda$CDM
 \be
 \eta=\frac{\rho_{by}}{\rho_{dm}+\rho_{by}},
 \en
 we immediately obtain from (\ref{rhodm})
 \be
 \mu=\eta.
 \en
  In MOND $\mu$ is not a certain function except the qualitative property that $\mu=1$ when $a/a_c>>1$ and $\mu=a/a_c$ when $a/a_c<<1$.
  However, it is faithful that the main implications for galaxies caused by MOND do not depend on the specific form of $\mu$. Here we adopt a fairly general form,
  {\large
  \be
  \mu=\frac{a/a_c}{\left[{1+(a/a_c)^{\alpha}}\right]^{\frac{1}{\alpha}}},
  \label{mu}
  \en}
  where $\alpha$ is a positive constant.
     Solving the above equation, we derive $a/a_c$,
   \be
   \frac{a}{a_c}=\frac{\eta}{\left({1-\eta^{\alpha}}\right)^{\frac{1}{\alpha}}}.
   \label{eta}
   \en
   According to the final result from WMAP+BAO+H0 Mean we have $\eta=1/6$ and $H_0=70.2$ kms$^{-1}$ Mpc$^{-1}$ \cite{wmap}.
   From (\ref{physicR}),
  \be
  \frac{a_{0}}{R_{0}}=-\frac{\ddot{R}_{0}r}{R_{0}}.
  \label{apr}
  \en
  Note that $a_0$ is the present value of $a$, which is different from $a_c$.
  We set the dimensionless comoving coordinate $r\in (0,1)$, and hence the scale factor $R$ has a dimension of length. $R$ is a geometric
  parameter, which is shared by frames of MOND cosmology and $\Lambda$CDM. We calculate $\frac{a_{0}}{R_{0}}$ in spatially flat $\Lambda$CDM.
  First,
  \be
   \frac{\ddot{R}_{0}r}{R_{0}}=-\left[\frac{4\pi G}{3}(\rho_{\rm tot}+3p_{\rm tot})\right]_{0}r,
   \label{Rr}
   \en

     \be
   H^2_{0}=\left[\frac{8\pi G}{3}\rho_{\rm tot}\right]_{0},
   \label{H2}
   \en
 where subscript tot denotes the total cosmic fluids.
  From (\ref{Rr}) and (\ref{H2}), we get
  \be
  \frac{\ddot{R}_{0}}{R_{0}}=-\left[\frac{ H^2}{2}\frac{\rho_{\rm tot}+3p_{\rm tot}}{\rho_{\rm tot}}\right]_{0},
  \label{ddr}
  \en
  in which we set $r=1$.  If we cancel the subscript 0 in the above equation, we obtain the general equation
  which is  tenable at any cosmic epoch.

  Then, from (\ref{eta}, \ref{apr}) and (\ref{ddr}), we obtain
  \be
  a_c=-q_0\frac{\left({1-\eta^{\alpha}}\right)^{\frac{1}{\alpha}}}{\eta}cH_{0},
  \label{acq}
  \en
  where $q$ denotes the deceleration parameter and we have used $R_{0}=H_{0}^{-1}$. In (\ref{acq}), we have restored all the dimensions by including the luminous velocity $c$. If we set $\alpha=3/10$, and borrow the numerical values of $H_{0}$, $q_0$, and $\eta$ from \cite{wmap}, we reach
  \be
  a_c=1.2\times 10^{-8} {\rm cm~ sec}^{-2}=0.176cH_0,
  \en
  which is well consistent with the value fitted by latest data of spiral galaxies \cite{spri}.
  Thus for the first time, we get the transition parameter $a_c$ in MOND from cosmological observations.

   Now we consider the virtual dark energy term (\ref{vde}). Cosmic acceleration is one of the most striking discoveries over last century \cite{acce}. MOND is suggested much earlier than the discovery of the cosmic acceleration. Of course, it is not designed to account for this acceleration. We shall show that (\ref{vde}) can accelerate the universe, therefore an accelerating universe can be treated as a prediction in a sense.

  We study the behavior of the virtual dark energy in a power law universe $R\sim t^n$. Any universe in a small time segment can be viewed as a power law universe. When $a/a_c<<1$, we get
  \be
  \rho_{de}=\frac{3}{8\pi G} \frac{n^2(n-2)}{3n-4}\frac{1}{t^2},
  \label{rholate}
  \en
  and when $a/a_c>>1$,
  \be
  \rho_{de}=0.
  \en
  So, $\rho_{de}$ becomes important when $a/a_c<<1$, i.e., in the late universe. Several notes on the dark energy (\ref{vde}) are listed as follows:

   1. The late behavior of dark energy in MOND cosmology (\ref{rholate}) also can be realized by K\'arolyh\'azy relation, which is an effect of quantum mechanics in general relativity \cite{cai2}. This kind of dark energy is called agegraphic dark energy in literatures.

   2. In the early time when $a/a_c>>1$, it disappears spontaneously such that it does not affect structure formation, though structure formation in MOND cosmology is still not clear.

   3. There is only one new parameter $a_c$ in this MOND cosmology compared to $\Lambda$CDM and it is no cosmological constant in MOND cosmology. Totally, it has the same freedoms as $\Lambda$CDM. Furthermore, there is really no free parameter in the dark energy term (\ref{vde}) (if we set a zero integration constant).

   4. This term appears in an integration form, which implies the evolution of the universe at any epoch depends on a period of the history of the universe. This non-localization property is a result of holographic principle.

  Now we make a preliminary numerical study in this model. As we have mentioned, any universe in a short time segment can be regarded as a power-law one. We use this point to estimate the present power $n$ and then the present deceleration parameter. Through comparing with $\Lambda$CDM, we reach
  \be
  \frac{8\pi G\rho_{de_{0}}}{3H_{0}^2}=0.725.
  \en
   Next, from (\ref{rholate}) we obtain $n=2.407$, which describes an accelerating universe whose deceleration parameter is $-0.5813$. The result perfectly matches recent observations. Note that we do not use the geometric information from the fitting results of $\Lambda$CDM, but only borrow the partition of dark energy density.  More serious results about the parameters in MOND cosmology need a global fitting by combining different observations.

   \section{conclusion}
   MOND is a very successful paradigm in galaxy scale to take the place of dark matter. We make a direct extension for MOND to the cosmological scale via EF method. We study the primary properties of this MOND cosmology in this letter. We find that MOND cosmology inherits the merits of the original MOND, that is, it can generate enough dust in view of standard model with only baryonic matter. Moreover, by means of the partition of dark matter and baryonic matter in $\Lambda$CDM we get the critical parameter $a_c$ in MOND on the cosmological scale for the first time. The other interesting property of MOND cosmology is that a dark energy  term which can accelerate the universe emerges spontaneously. We make a preliminary investigation about this term and find that it perfectly consistent  with observations. Several details about this model need to explore further.

 {\bf Acknowledgments.}
  This work is supported by the Program for Professor of Special Appointment (Eastern Scholar) at Shanghai Institutions of Higher Learning, National Education Foundation of China under grant No. 200931271104, Shanghai Municipal Pujiang grant No. 10PJ1408100, and National Natural Science Foundation of China under Grant No. 11075106.


\begin{thebibliography}{99}

 \bibitem{darkmatter}
 F. Zwicky,
 %"Die Rotverschiebung von extragalaktischen Nebeln".
  Helvetica Physica Acta 6: 110-127(1933);P.~G.~Ferreira and G.~Starkmann,
  %``Einstein's Theory of Gravity and the Problem of Missing Mass,''
  Science {\bf 326}, 812 (2009).
  \bibitem{mil1}
  M. Milgrom,  ApJ 270, 365(1983).

  \bibitem{revmond}
 M.~Milgrom,
  %``The MOND paradigm,''
  arXiv:0801.3133 [astro-ph]; J.~D.~Bekenstein,
  %``Relativistic MOND as an alternative to the dark matter paradigm,''
  Nucl.\ Phys.\  A {\bf 827}, 555C (2009)
  ; %``Alternatives to dark matter: Modified gravity as an alternative to dark
  %matter,''
  arXiv:1001.3876 [astro-ph.CO].

 \bibitem{tully}
  E. Opik, Astrophys. J., 55, 406(1922);
 R. Tully , J. Fisher, Astron. Astrophys., 54, 661(1977).

 \bibitem{mc}
  R.~H.~Sanders and S.~S.~McGaugh,
  %``Modified Newtonian dynamics as an alternative to dark matter,''
  Ann.\ Rev.\ Astron.\ Astrophys.\  {\bf 40}, 263 (2002)
  [arXiv:astro-ph/0204521].
  %%CITATION = ARAAA,40,263;%%


  \bibitem{beken1}
  J. Bekenstein, Phys.Rev. D70, 083509(2004).
  \bibitem{hao1}
  J. Hao, R. Akhoury, Int. J. Mod. Phys. D18:1039(2009).
\bibitem{ver}
  E.~P.~Verlinde,
  %``On the Origin of Gravity and the Laws of Newton,''  JHEP {\bf 1104}, 029 (2011)  [arXiv:1001.0785 [hep-th]].

  \bibitem{shu}
 F-W, Shu and Y. Gong, %¡°Equipartition of energy and the
 %first law of thermodynamics at the apparent horizon,¡±
arXiv:1001.3237 [gr-qc].
\bibitem{cai1}
R-G, Cai, L-M cao and N. Ohta, %¡°Friedmann Equations
%from Entropic Force,¡±
Phys. Rev. D 81, 061501(2010).

  \bibitem{gao}
  C.~Gao,
  %``Modified Entropic Force,''
  Phys.\ Rev.\  D {\bf 81}, 087306 (2010)
  [arXiv:1001.4585 [hep-th]];H.~Wei,
  %``Cosmological Constraints on the Modified Entropic Force Model,''
   Phys.\ Lett.\ B {\bf 692}, 167 (2010)  [arXiv:1005.1445 [gr-qc]].
  \bibitem{saha}
  A.~D.~Sakharov,
%``Vacuum quantum fluctuations in curved space and the theory of gravitation,''
Sov.\ Phys.\ Dokl.\  {\bf 12}, 1040 (1968).

\bibitem{lyx}
  S.~-W.~Wei, Y.~-X.~Liu and Y.~-Q.~Wang,
  %``A note on Friedmann equation of FRW universe in deformed Horava-Lifshitz gravity from entropic force,''
   Commun.\ Theor.\ Phys.\  {\bf 56}, 455 (2011)  [arXiv:1001.5238 [hep-th]].  %%CITATION = ARXIV:1001.5238;%%

   \bibitem{lingyi}
  Y.~Ling and J.~-P.~Wu,
  %``A note on entropic force and brane cosmology,''
  JCAP {\bf 1008}, 017 (2010)  [arXiv:1001.5324 [hep-th]].

  \bibitem{chiba}
   N.~Shiba,
  %``Entanglement Entropy of Two Black Holes and Entanglement Entropic Force,''
  Phys.\ Rev.\ D {\bf 83}, 065002 (2011)  [arXiv:1011.3760 [hep-th]].

  \bibitem{jaco}
  T. Jacobson,  Phys.
Rev. Lett. 75, 1260 (1995) .

  \bibitem{pad}
  T. Padmanabhan,  Class. Quant. Grav. 19, 5387 (2002) [arXiv:gr-qc/0204019].
  \bibitem{feng1}
  C.~J.~Feng, X.~Z.~Li and X.~Y.~Shen,
  %``Thermodynamic of the QCD Ghost Dark Energy Universe,''
  arXiv:1105.3253 [hep-th]; arXiv:1101.5843 [hep-th].

   \bibitem{unru}
    W.~G.~Unruh,
  %``Notes on black hole evaporation,''
  Phys.\ Rev.\  D {\bf 14}, 870 (1976).
  %%CITATION = PHRVA,D14,870;%%
  \bibitem{lyx2}
  Y.~-X.~Liu, Y.~-Q.~Wang and S.~-W.~Wei,
  %``A Note on Temperature and Energy of 4-dimensional Black Holes from Entropic Force,''
  Class.\ Quant.\ Grav.\  {\bf 27}, 185002 (2010)  [arXiv:1002.1062 [hep-th]]; E.~Chang-Young, M.~Eune, K.~Kimm and D.~Lee,
  %``Surface gravity and Hawking temperature from entropic force viewpoint,''
  Mod.\ Phys.\ Lett.\ A {\bf 25}, 2825 (2010)  [arXiv:1003.2049 [gr-qc]].

 \bibitem{reviewcross}
  H Zhang, Crossing the phantom divide, in $ Dark~ Energy:~ Theories,~ Developments~ and~ Implications$,
  Nova Science Publisher (2010).
  \bibitem{wmap}
  E.~Komatsu {\it et al.}  [WMAP Collaboration],
  %``Seven-Year Wilkinson Microwave Anisotropy Probe (WMAP) Observations:
  %Cosmological Interpretation,''
  Astrophys.\ J.\ Suppl.\  {\bf 192}, 18 (2011)
  .
 \bibitem{spri}
  S.~S.~McGaugh,
  %``A Novel Test of the Modified Newtonian Dynamics with Gas Rich Galaxies,''
  Phys.\ Rev.\ Lett.\  {\bf 106}, 121303 (2011)
  [arXiv:1102.3913 [astro-ph.CO]].

  %%CITATION = APJSA,192,18;%%
  \bibitem{acce}
  A. G. Riess et al.,
  Astron. J. 116, 1009 (1998);
  S. Perlmutter et al.,
  Astrophys. J. 517, 565 (1999).
  \bibitem{cai2}
  R.~G.~Cai,
  %``A Dark Energy Model Characterized by the Age of the Universe,''
  Phys.\ Lett.\  B {\bf 657}, 228 (2007).











\end{thebibliography}
\end{document}